\documentstyle[aps]{revtex}
\begin{document}
\draft
\title{New integrable model of correlated electrons with off-diagonal
long-range order from $so(5)$ symmetry }
\author{Angela Foerster,$^1$ Jon Links$^{1,2}$ and Itzhak Roditi$^{1,3}$ }

\address{ $^1$Instituto de F\'{\i}sica da UFRGS, Av. Bento Gon\c{c}alves
9500, Porto Alegre, RS - Brazil }

\address{$^2$Department of Mathematics, University of
Queensland,  4072,  Australia }

\address{$^3$Centro Brasileiro de
Pesquisas F\'{\i}sicas - CBPF,
Rua Dr. Xavier Sigaud 150, 22290-180, Rio de Janeiro, RJ - Brazil }

\maketitle
\begin{abstract}
We present a new integrable model for correlated electrons which
is based on a $so(5)$ symmetry. By using an $\eta$-pairing
realization we construct eigenstates of the Hamiltonian with
off-diagonal long-range order. It is also shown that these
states lie in the ground state sector. We exactly solve the model
on a one-dimensional lattice by the Bethe ansatz.  
\end{abstract}
\vspace{0.35cm}
The study of models of correlated electrons is a significant tool in the
theory of condensed matter physics. On a one
dimensional lattice there are several known models which are
exactly solvable by Bethe ansatz methods. The most famous of these is
the Hubbard model whose solution was obtained by Lieb and Wu \cite{li}.
Another well known example is the $t-J$ model, the strong-coupling
limit of the Hubbard model, which was in fact shown to be integrable
at the supersymmetric point \cite{fab1} through use of the Quantum Inverse
Scattering Method (QISM) \cite{skly}. In this formalism the Hamiltonian of the
model is derived from a solution of the Yang-Baxter equation, hereafter
referred to as an $R$-matrix,  which provides
a systematic method to obtain higher order conservation laws that guarantee
integrability. An important aspect of the integrable coupling of the $t-J$
model is that the $R$-matrix is invariant with respect to
the Lie superalgebra $gl(2|1)$.
For the case of the Hubbard model the symmetry algebra has been
identified as $so(4)$ \cite{fab2}.

A further important integrable correlated electron model was proposed
and solved through the algebraic Bethe ansatz method by
Essler et. al. \cite{fab3}. This model
generalizes the Hubbard model with the addition of correlated hopping
and pair hopping terms and is constructed from an $R$-matrix invariant with
respect to the Lie superalgebra $gl(2|2)$. Another direction of generalization
was given by Bracken et. al. \cite{j1} using the $R$-matrix obtained from the
one-parameter family of four dimensional representations of $gl(2|1)$. The
resulting model, known as the supersymmetric $U$ model has also been solved
and analyzed by Bethe ansatz techniques \cite{j2}.
In all the above examples the underlying symmetry has crucial consequences
for the multiplet structure of the models providing insight into the
ground state and elementary excitations.

Recently it has been proposed that the antiferromagnetic and
superconducting phases of high-$T_c$ cuprate compounds are
unified by an approximate $so(5)$ symmetry \cite{sc1}.
Considerable support for this proposal came from numerical
investigations in models for high-$T_c$ materials.
In particular, it was shown that
the low-energy excitations can
be classified in terms of an $so(5)$ symmetry multiplet
structure \cite{sc2}.
Subsequently extended Hubbard models related with an $so(5)$ symmetry
have been introduced and analysed in detail \cite{hen}.

To our knowledge, no integrable correlated electron
model associated with an $so(5)$ symmetry has been proposed nor exactly
solved.
In this paper we construct such a correlated electron model
which is exactly solved in one dimension by
the Bethe ansatz.
The integrability of our Hamiltonian comes from the
fact that it is derived from an $so(5)$ invariant $R$-matrix
which satisfies the quantum Yang-Baxter equation.
Eigenstates of this Hamiltonian
exhibiting off-diagonal long-range order (ODLRO) can be constructed
through an $\eta$-pairing mechanism. We also argue
that these states lie in the ground state sector,
which is a prerequisite for superconductivity.

The Hamiltonian of this model is given by
\begin{equation}
\label{1}
H=\sum_{i=1}^{L-1} h_{i,i+1} + h_{L,1} + \mu N + B S^z
\end{equation}
where
\begin{eqnarray}
h_{i,j} &=& - \sum_{\sigma= \uparrow, \downarrow}  c^{\dagger}_{i \sigma} c_{j\sigma}
( 3 - 2 n_{i, -\sigma} -4 n_{j, -\sigma} ) + h.c.
\nonumber \\
&-& 2( c^{\dagger}_{i\uparrow} c^{\dagger}_{i\downarrow}
c_{j\downarrow} c_{j\uparrow}  + h.c. )
- 4 ( S_i^x S_{j}^x + S_i^y S_{j}^y + 3  S_i^z S_{j}^z )
\nonumber \\
&-& 3(n_i + n_{j}) - 3 n_i n_{j} -
4 [( 1/4 (n_i - n_j)^2 - (S_i^z - S_j^z)^2 ]^2
\label{2}
\end{eqnarray}
Above $c_{i\sigma}, \, c_{i\sigma}^{\dagger}$ are annihilation and
creation operators for electrons of spin $\sigma$, the $\vec S_i$'s
spin matrices and the $n_{i \uparrow},\,n_{i\downarrow}$'s
occupation numbers of electrons at
lattice site $i$. The number of lattice sites is $L$,
$S^z = \sum_{i=1}^L S^z_i$ is the magnetization and
$N = \sum_{i=1}^{L} (n_{i \uparrow}  + n_{i \downarrow})$
is the number of electrons.
This Hamiltonian exhibits correlated electron hoppings,
pair hoppings, XXZ type interaction, chemical potential,
nearest neighbouring Coulomb interaction, and the last term characterizes
interactions favouring antiferromagnetism.
The energy levels of the model are
\begin{equation}
\label{3}
E = \sum_j { 1 \over 4 u_j^2 - 1} + \mu N + B S^z
\end{equation}
where the $u_j$'s are solutions of the Bethe ansatz equations
\begin{eqnarray}
{\biggl(
{ u_i + 1/2 \over u_i - 1/2 }  \biggr)}^L &=&
- (-1)^{M_1}
\prod_{j \neq i}^{M_1} {u_i - u_j + 1 \over u_i - u_j - 1}
\prod_{k}^{M_2} {u_i - \overline{u_k} - 1 \over u_i - \overline{u_k} + 1},
\, \,
i = 1, \dots M_1  , \nonumber \\
1 &=&
\prod_{j}^{M_1} {u_j - \overline{u_i} - 1 \over u_j - \overline{u_i} + 1}
\prod_{k \neq i}^{M_2} {\overline{u_k} - \overline{u_i} + 2 \over
\overline{u_k} - \overline{u_i} - 2},
\, \,
j = 1, \dots M_2 ,
\label{4}
\end{eqnarray}
where $M_1 = 2L - N$ and $M_2 = L - N_{\uparrow}$.
Integrability of this model will be established through the
QISM. The energy eigenvalues as well as the Bethe ansatz
equations are obtained through the analytic
Bethe ansatz \cite{re}. The key ingredient to both of these methods is
the following $R$-matrix \cite{mar1}
\begin{equation}
R(u) = \sum_{i,j} \biggl( u(u-3)e^i_i \otimes e^j_j +
(3-u) e^i_j \otimes e^j_i + (-1)^{i+j} u \, e^i_j \otimes
e^{\overline{i}}_{\overline{j}} \biggr)
\label{5}
\end{equation}
which satisfies the Yang-Baxter equation
\begin{equation}
R_{12}(u-v)R_{13}(u)R_{23}(v)=R_{23}(v)R_{13}(u)R_{12}(u-v).
\label{ybe}
\end{equation}
Above the matrices $e^i_j$ have entries $(e^i_j)^k_l = \delta_{ik}\delta_{jl}$,
the indices range from 1 to 4 and $\overline{i} = 5-i$.
This $R$ matrix possesses the properties of
\begin{itemize}
\item [--] Unitarity
\begin{equation}
R(u) R(-u) = (u^2-1) (u^2-9)I \otimes I
\label{6}
\end{equation}
\item [--] Crossing-symmetry
\begin{equation}
R^{t_1}(u) = - A_1 R(3-u) A_1
\label{7}
\end{equation}
\end{itemize}
where $t_1$ denotes transposition in the first space and
$A = \sum_i (-1)^i e^i_{\overline i}$.

The solution (\ref{5}) is invariant with respect to the Lie algebra $ so(5)
\cong sp(4)$ which has ten generators
\begin{equation}
a^i_j = e^i_j-(-1)^{i+j}e^{\overline j}_{\overline i}
= -(-1)^{i+j} a^{\overline j}_{\overline i}
\label{8}
\end{equation}
satisfying the commutation relations
$$[a^i_j,\,a^k_l]=\delta^k_j a^i_l-\delta^i_l a^k_j+(-1)^{i+j}
\delta^{\overline j}_l a^k_{\overline i}-(-1)^{i+j}
\delta^k_{\overline i} a^{\overline j}_l.    $$

In order to build an electronic model we first need to
put ${\bf Z}_2$ grading in the $R$-matrix. This is achieved by
a redefinition of the matrix elements through
\begin{equation}
{R(u)}^{i j}_{k l} \rightarrow (-1)^{[i][j]+[k][j]+[k][l]} {R(u)}^{i j}_{k l}
\label{9}
\end{equation}
where we choose the parities to be
$$ [1] = [4] = 0, \, \, \, \, \, \, \, \,  [2] = [3] = 1
\, \, \, \, \, \, \, \, \, \, \, \,  {\rm and}
\, \, \, \, \, \, \, \,
[e^i_j] = [i] + [j]. $$
Such a matrix satisfies the ${\bf Z}_2$ graded Yang-Baxter
where the multiplication of tensor products of matrices is governed by
$$ (a \otimes b) (c \otimes d) = (-1)^{[b][c]} ac \otimes bd $$
in eq. (\ref{ybe}).
Following the QISM, we may construct the transfer matrix
\begin{equation}
\tau(u) = str_0 \biggl( R_{0L}(u) R_{0 L-1}(u) \dots R_{02}(u)
R_{01}(u) \biggr),
\label{tr}
\end{equation}
where $str_0$ is the supertrace over the zeroth space. From
the Yang-Baxter algebra it follows that the transfer matrices
$\tau(u)$ form a commuting family and the associated Hamiltonian (\ref{1})
with $\mu=0$ and $B=0$ can be obtained from
$$ H = { \tau(u)}^{-1} \left. \frac{d}{du} \tau(u) \right|_{u=0}, $$
where in view of the grading we have used the following identification
$$ \left |1 \right> \equiv \left | \uparrow \downarrow \right>, \, \, \, \,
   \left |2 \right> \equiv \left | \uparrow \right>, \, \, \, \, \,
   \left |3 \right> \equiv \left | \downarrow \right>, \, \, \, \, \,
   \left |4 \right> \equiv \left | 0 \right> \,
   .
   \nonumber $$
In terms of the fermion operators, the $so(5)$ generators (\ref{8})
can be written as
\begin{eqnarray}
a^1_1 &=& n - 1, \, \, \,
a^2_2 = 2 S^z, \, \, \,
a^1_2 = c^\dagger_-, \, \, \,
a^2_1 = c_-, \, \, \,
a^1_3 = -c^\dagger_+, \nonumber \\
a^3_1 &=& -c_+, \, \, \,
a^1_4 = 2 c^\dagger_- c^\dagger_+, \, \, \,
a^4_1 = 2 c_+ c_-, \, \, \,
a^2_3 = 2 S^+, \, \, \,
a^3_2 = 2 S^-.
\label{10}
\end{eqnarray}
On the 2-fold tensor product space these generators act according to
the co-product
\begin{eqnarray}
\Delta(a^i_j)&=& a^i_j \otimes I + (-1)^n \otimes a^i_j\, \, \,\, \, \, \,
 {\rm for} \, \, \, \, \, \, \, a^i_j = a^1_2, a^2_1, a^1_3, a^3_1,
 \nonumber \\
\Delta(a^i_j)&=& a^i_j \otimes I + I \otimes a^i_j\, \, \,\, \, \, \,
 {\rm otherwise},
\label{11}
\end{eqnarray}
which extends to the $L$-fold tensor space co-associatively.
Each of the local Hamiltonians $h_{i,i+1}$ (\ref{2}) are
$so(5)$ invariant. However due to the
non-cocommutativity of the co-product the $h_{L,1}$ term
breaks the $so(5)$ symmetry of the global Hamiltonian (\ref{1}).
In spite of this, an $so(4)$ symmetry is preserved comprising of
an $so(3)$ spin realization and an additional $so(3) \,  \eta $
pairing realization. For this reason we can add arbitrary
chemical potential and magnetic field terms to the Hamiltonian
which do not violate the integrability.

The presence of the $\eta$ pairing realization
\begin{eqnarray}
\eta =\sum_{j=1}^L c_{j,\uparrow} c_{j,\downarrow}\,,~~\eta^{\dagger}=
\sum_{j=1}^L
c^{\dagger}_{j,\downarrow} c^{\dagger}_{j,\uparrow}\,,~~ \eta^z=
\sum_{j=1}^L \frac 12 (n_j-1).
\label{12}
\end{eqnarray}
which can also be expressed in terms of the
subalgebra generated by $\{a^1_1, a^1_4, a^4_1\}$, allows
a large number of states to be constructed exhibiting
ODLRO \cite{yang}. Hereafter we treat the case $\mu=0, \, B=0$.
One can verify that
$H\left|0\right>=0$ where $\left|0\right>$ denotes
the vacuum state.
Thus the $2{\cal N}$ electron states
\begin{equation}
\left|\Psi_{\cal N}\right> =(\eta^{\dagger})^{\cal N}\left|0\right>
\label{13}
\end{equation}
are eigenstates of the global Hamiltonian with zero energy.  These states are
well known to possess ODLRO; that is
\begin{eqnarray}
\lim_{|l-j|\rightarrow\infty} \frac{\left<\Psi_{\cal N}\right|
c^{\dagger}_{j,\downarrow}c^{\dagger}_{j,\uparrow}c_{l,\uparrow}
c_{l,\downarrow}\left|\Psi_{\cal N}\right>}{\left<\Psi_{\cal N}|
\Psi_{\cal N}\right>}~~~
=\frac{ {\cal{ N}}}{L} \left( 1-\frac{{\cal N}}{L}\right)
\end{eqnarray}
in the thermodynamic limit (${\cal N},L\rightarrow \infty,~~ {\cal N}/L$
fixed).
Since the Hamiltonian is Hermitian the ground state
energy satisfies
$$E\geq LE_0$$
where $E_0$ is the minimum energy of the two-site Hamiltonian.
For this model we can determine that $E_0 = 0$.
It is thus concluded that the states (\ref{13}) lie in the
ground state sector.

The energy levels (\ref{3}) are determined from the
eigenvalues of the transfer matrix (\ref{tr})
which leads to a complicated expression that we
will not give here. However, we mention that
these eigenvalues are obtained through the analytic
Bethe ansatz which exploits the properties of
unitarity (\ref{6}), crossing symmetry (\ref{7})
and asymptotic behaviour of the $R$-matrix.
As usual, the Bethe ansatz equations are
derived by the requirement that the eigenvalues
are analytic functions.

In conclusion, we have introduced a new integrable
correlated electron model based on a $so(5)$ symmetry.
The model was exactly solved through the Bethe ansatz
and shown to have ground states exhibiting ODLRO.

\vspace{0.35cm}

The authors thank Funda\c{c}\~ao de Amparo a Pesquisa
do Estado do Rio Grande do Sul, Conselho Nacional de Desenvolvimento
Cient\'{\i}fico e Tecnol\'ogico and Australian Research Council
for financial support.
J. L. and I. R. thank the Instituto de F\'{\i}sica da UFRGS for
their kind hospitality.


\end{document}